\documentclass{PoS}
\usepackage{psfrag,epsfig,graphicx,graphics}

\usepackage{mathptmx}  

\usepackage{wrapfig}
\usepackage{epsf}
\usepackage{amssymb}
\usepackage{amsmath}
\usepackage{amsfonts}

\def\lsim{\raise0.3ex\hbox{$<$\kern-0.75em\raise-1.1ex\hbox{$\sim$}}}
\def\gsim{\raise0.3ex\hbox{$>$\kern-0.75em\raise-1.1ex\hbox{$\sim$}}}

\newcommand{\half}{ {\textstyle\frac{1}{2}} }

\newcommand{\tvec}[1]{{{#1}_\perp}}

\newcommand{\rb}{\underline{r}}
\newcommand{\kb}{\underline{k}}

\newcommand{\odd}{\mathbb{O}}
\newcommand{\pom}{\mathbb{P}}

\def\bei{\begin{itemize}}
\def\ei{\end{itemize}}

\def\beeq{\begin{eqnarray}} 
\def\beqa{\begin{eqnarray}}
\def\bea{\begin{eqnarray}}

\def\eea{\end{eqnarray}}
\def\eqa{\end{eqnarray}}
\def\eeeq{\end{eqnarray}}

\def\beas{\begin{eqnarray*}}
\def\beqas{\begin{eqnarray*}}

\def\eqas{\end{eqnarray*}}
\def\eeas{\end{eqnarray*}}

\def\beq{\begin{equation}} 
\def\be{\begin{equation}}

\def\ee{\end{equation}}
\def\eq{\end{equation}}
\def\eeq{\end{equation}}

\def\beqd{\begin{displaymath}}
\def\eeqd{\end{displaymath}}
\def\eqd{\end{displaymath}}

\def\beeq{\begin{eqnarray}} \def\eeeq{\end{eqnarray}}

\newcommand{\fin}{\end{document}}

\def\bef{\begin{frame}}

\def\aut{\usebeamercolor[fg]{example text}}
\def\stmath{\usebeamercolor[fg]{structure}}

\def\stmath{}
\def\alert{}
\def\aut{}
\def\structure{}
\def\twist{}
\def\twist3{}

\def\reduced{\scriptsize}

\newcommand{\widm}{0.5\columnwidth}

\def\slashchar#1{\setbox0=\hbox{$#1$}
   \dimen0=\wd0
   \setbox1=\hbox{/} \dimen1=\wd1
   \ifdim\dimen0>\dimen1
      \rlap{\hbox to \dimen0{\hfil/\hfil}}
      #1
   \else
      \rlap{\hbox to \dimen1{\hfil$#1$\hfil}}
      /
   \fi}

\def\bei{\begin{itemize}}
\def\ei{\end{itemize}}

\def\structure#1{#1}
\def\stmath#1{#1}
\def\aut#1{#1}
\def\alert#1{#1}

\title{Exclusive Processes: Theory Introduction}

\ShortTitle{Exclusive Processes: Theory Introduction}

\author{{Samuel Wallon}\\
       LPT, Universit\'e Paris-Sud, CNRS, 91405, Orsay, France\\
       \&\\
       UPMC Univ. Paris 06, facult\'e de physique, 4 place Jussieu, 75252 Paris Cedex 05,
France\\
       E-mail: \email{Samuel.Wallon@th.u-psud.fr}}

\abstract{
We review the recent developments on the theoretical description of exclusive processes at medium and asymptotical energies. These are illustrated based on a few examples.}

\FullConference{Photon 2013,\\
		20-24 May 2013\\
		Paris, France }

\begin{document}

\section{Introduction}

\subsection{Exclusive processes in the early days of QCD}
\label{subsec:early}

\begin{figure}[h]
\centerline{\begin{tabular}{cc}
\psfrag{pq}{}
\psfrag{pg1}{}
\psfrag{pqb}{}
\psfrag{p1}{}
\psfrag{pp1}{}
\psfrag{p2}{}
\psfrag{pp2}{}
\psfrag{pu1}{}
\psfrag{pd1}{}
\psfrag{ppu1}{}
\psfrag{ppd1}{}
\psfrag{pu2}{}
\psfrag{pd2}{}
\psfrag{ppu2}{}
\psfrag{ppd2}{}
\psfrag{k}{}
\psfrag{kp}{}
\hspace{1cm}\includegraphics[height=2.5cm]{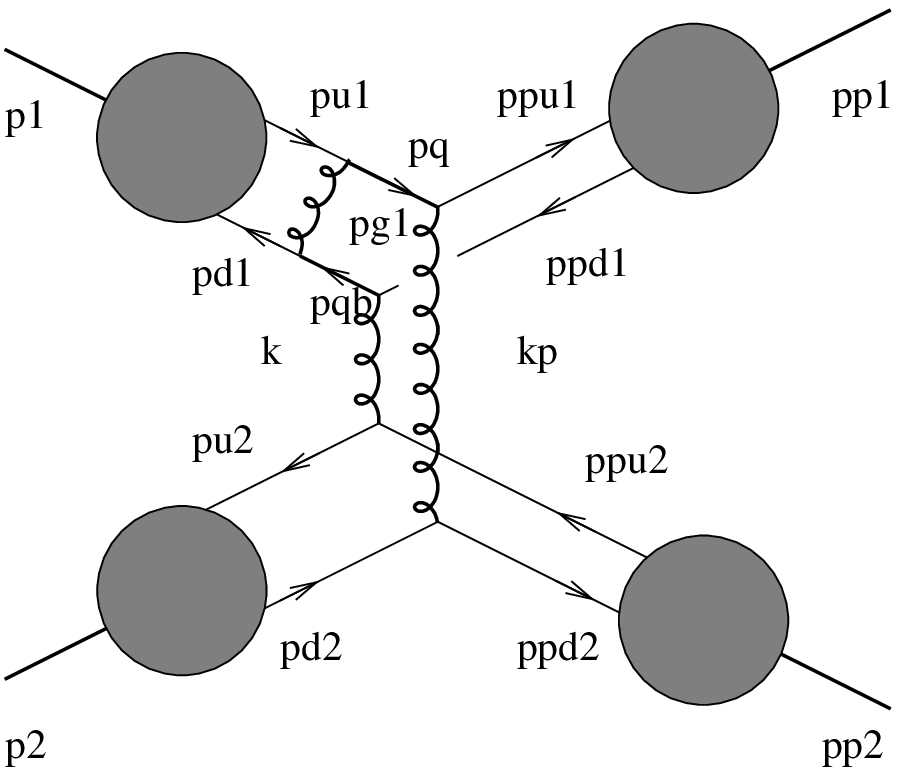} \quad \raisebox{1cm}{(a)}& 
\psfrag{pq}{}
\psfrag{pg1}{}
\psfrag{pqb}{}
\psfrag{p1}{}
\psfrag{pp1}{}
\psfrag{p2}{}
\psfrag{pp2}{}
\psfrag{pu1}{}
\psfrag{pd1}{}
\psfrag{ppu1}{}
\psfrag{ppd1}{}
\psfrag{pu2}{}
\psfrag{pd2}{}
\psfrag{ppu2}{}
\psfrag{ppd2}{}
\psfrag{k}{}
\psfrag{kp}{}
\hspace{3cm}\includegraphics[height=2.5cm]{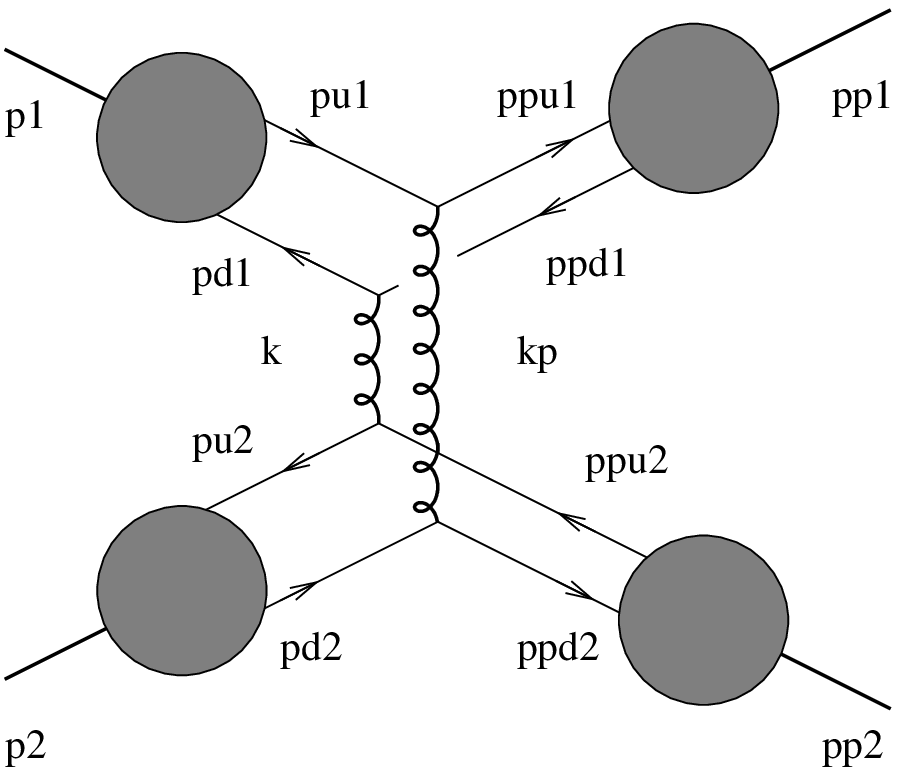} \quad \raisebox{1cm}{(b)}
\end{tabular}}
\caption{Large angle $\pi \pi \to \pi \pi$ scattering. Brodsky-Lepage (a) and  Landshoff (b) mechanisms.}
\label{Fig:BL-Landshoff}
\end{figure}

In recent years, hard
exclusive processes proved to be very efficient tools in order to get insight 
into the internal tri-dimensional partonic structure of hadrons.
The main question is whether one can 
extract information on hadrons using \alert{hard} exclusive processes, in a reliable way.
The aim is to reduce 
the process to interactions involving a small number of {\it partons}
(quarks, gluons), despite confinement.
\begin{figure}[h!]
\scalebox{1.1}{\begin{tabular}{cc}
\psfrag{k}{\hspace{-.3cm}$e^-$}
\psfrag{kp}{\hspace{-.1cm}$e^-$}
\psfrag{g}{\raisebox{-.2cm}{$\hspace{-.4cm}\gamma^*$}}
\psfrag{P}{$p$}
\psfrag{Q}{\raisebox{-.2cm}{$p$}}
\psfrag{q}{$q$}
\psfrag{p}{}
\psfrag{ppq}{}
\psfrag{pro}{\scalebox{.7}{\hspace{-.5cm}\structure{hard partonic process}}}
\hspace{.4cm}\includegraphics[height=2.2cm]{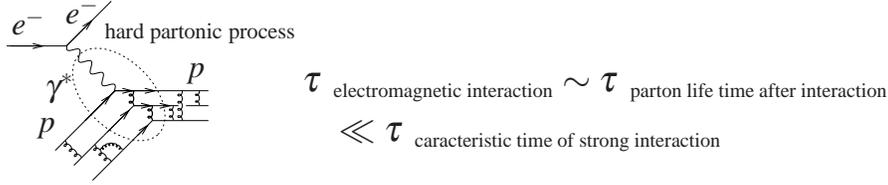} 
&
\hspace{.5cm}\raisebox{.8cm}{\scalebox{1.15}{
$\begin{array}{l}\tau_{\mbox{\, \tiny electromagnetic interaction}} \sim
\tau_{\mbox{\, \tiny parton life time after interaction}}\\
 \quad \ll \tau_{\mbox{\, \tiny caracteristic time of strong interaction}}
\end{array}$}}
\end{tabular}}
\caption{Hard subprocess for the proton form factor, with the typical time scales involved.}
\label{Fig:partonsFormFactor}
\end{figure}
This is possible if the considered process is driven by short distance phenomena, allowing the use of perturbative methods.
One should thus hit strongly enough a hadron.
 This is typically what occurs in the case of 
an electromagnetic probe, which gives access to form factors, as illustrated in Fig.~\ref{Fig:partonsFormFactor}.
In practice, exploiting
such situations in exclusive reactions is very challenging since the  cross section are very small.
This weakness can be quantified based on counting rules, which shows that~\cite{Brodsky:1973krBrodsky:1974vy}
\beq
 F_n(q^2) \simeq \frac{C}{(Q^2)^{n-1}}
\label{countingFn}
\eq
where $n$ is the minimal number of constituents (meson: $n=2$; baryons: $n=3$). 
This result can be easily proven by considering the hard subprocess and evaluating the dimensions of the hard quark and gluons propagators, as well as the dimension of the $n$ quasi-free collinear quark degrees of freedom. 
A similar counting rule can be proven for 
\alert{large angle} (i.e. \structure{$s \sim t \sim u$ large}) elastic processes $h_a \, h_b \to h_a \, h_b$, e.g. 
\structure{$\pi \pi \to \pi \pi$}
or \structure{$p \, p \to p \, p$}, leading to~\cite{Brodsky:1981rp}
\beqa
\frac{d \sigma}{dt} \sim \left(\frac{\alpha_S(p_\perp^2)}s  \right)^{n-2}\, 
\label{scaling-large-angleBL}
\eqa
where
$n$ is the number of external fermionic lines
 ($n=8 \hbox{ for }  \pi \pi \to \pi \pi$). 
Limitations to the underlying factorized description have been known since decades, 
since other contributions might be significant, even at large angle~\cite{Landshoff:1974ew}. Consider for example
the process \structure{$\pi \pi \to \pi \pi$}.
The first mechanism (see Fig.~\ref{Fig:BL-Landshoff}a) 
relies on the description of each mesons through their collinear $q \bar{q}$ 
content, 
which longitudinal component along each meson momentum is 
encoded in their distribution amplitudes (DA), the whole amplitude scaling like
$\frac{d\sigma_{BL}}{dt} \sim  s^{-6}$
(see Eq.~(\ref{scaling-large-angleBL})). On the other hand, a competing mechanism may exist, with so-called pinched loop contributions (Fig.~\ref{Fig:BL-Landshoff}b). It assumes that particular collinear quark configurations of non-perturbative origin are present inside each meson. 
Thus, the additional hard gluon required to force the $q \, \bar{q}$ pair to be collinear in the mechanism of Fig.~\ref{Fig:BL-Landshoff}a is absent in figure  Fig.~\ref{Fig:BL-Landshoff}b, leading to a disconnected hard part. 
This contribution leads to a scaling
$\frac{d\sigma_{L}}{dt} \sim  s^{-5}\,.$ Note that this second mechanism is absent when at least one $\gamma^{(*)}$ is involved, due to its point-like coupling enforcing the presence of an additional gluon as in Fig.~\ref{Fig:BL-Landshoff}a.

\subsection{Recent experimental and theoretical developments}

The main difference between inclusive and exclusive processes is the hard scale power suppression, making the measurements much
more involved. This requires
high luminosity accelerators and high-performance detection facilities, as provided by
HERA (H1, ZEUS), HERMES, JLab@6 GeV (Hall A, CLAS), BaBar, Belle, BEPC-II (BES-III), LHC. Future projects will be essential for that purpose (COMPASS-II, JLab@12 GeV, LHeC, EIC, ILC).

In parallel to experimental developments, theoretical efforts have been 
very important  during the last decade... and many new acronyms for non-perturbative quantities have been popularized: DAs (distribution amplitudes), GPDs (generalised parton distributions), GDAs (generalized distribution amplitudes), TDAs (transition distribution amplitudes) and TMDs (transverse 
momentum dependent distributions), which we will try to introduce in a nutshell\footnote{For reviews, see~\cite{Guichon:1998xvGoeke:2001tz, Diehl:2003ny, Belitsky:2005qn, Boffi:2007ycBurkert:2007zzGuidal:2008zzaWallon:2011zx}.}.
These make sense in a given factorization framework, either at medium energies (collinear factorization) or asymptotical energies ($k_T$-factorization), allowing to 
deal both with perturbative and power corrections.



\section{Collinear factorizations}

\subsection{From DIS to exclusive processes}

\begin{figure}[h!]
\scalebox{.88}{\centerline{\hspace{1.2cm}\begin{tabular}{cc}
\raisebox{-.44 \totalheight}
{\psfrag{ph1}{$\gamma^*$}
\psfrag{ph2}{$\gamma^*$}
\psfrag{s}{$s$}
\psfrag{t}{$t$}
\psfrag{hi}{$p$}
\psfrag{hf}{$p$}
\psfrag{Q1}{\raisebox{-.2cm}{\,\,$Q^2$}}
\psfrag{Q2}{\raisebox{-.2cm}{\hspace{-.3cm}$Q^2$}}
\psfrag{x1}{$x$}
\psfrag{x2}{$x$}
\psfrag{GPD}{\raisebox{-.05cm}{\hspace{-.1cm}PDF}}
\psfrag{CF}{\raisebox{-.05cm}{\hspace{-.1cm}CF}}
\hspace{0cm}\scalebox{.8}{\raisebox{0 \totalheight}{\hspace{2cm}\includegraphics[height=5cm]{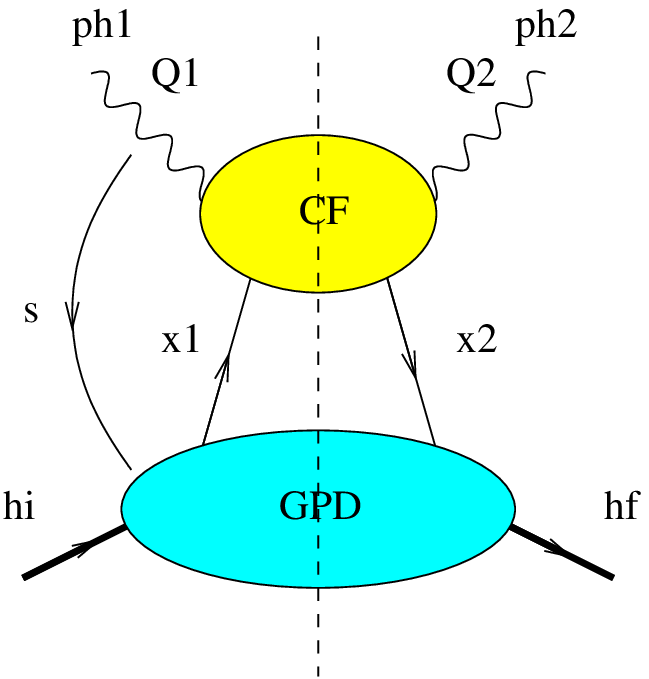}}}}
\raisebox{-.44 \totalheight}
{\psfrag{ph1}{$\gamma^* \, [\gamma]$}
\psfrag{ph2}{$\gamma\, [\gamma^*]$}
\psfrag{s}{$s$}
\psfrag{t}{$t$}
\psfrag{hi}{$p$}
\psfrag{hf}{$p'$}
\psfrag{Q1}{\raisebox{-.2cm}{\,\,$Q^2$}}
\psfrag{Q2}{\raisebox{-.2cm}{\hspace{-.5cm}$[Q^2]$}}
\psfrag{GPD}{\raisebox{-.05cm}{\hspace{-.1cm}GPD}}
\psfrag{CF}{\raisebox{-.05cm}{\hspace{-.1cm}CF}}
\psfrag{x1}{\hspace{-.4cm}$x+\xi$}
\psfrag{x2}{$x-\xi$}
\hspace{2.5cm}\scalebox{.8}{\raisebox{0 \totalheight}{\includegraphics[height=5cm]{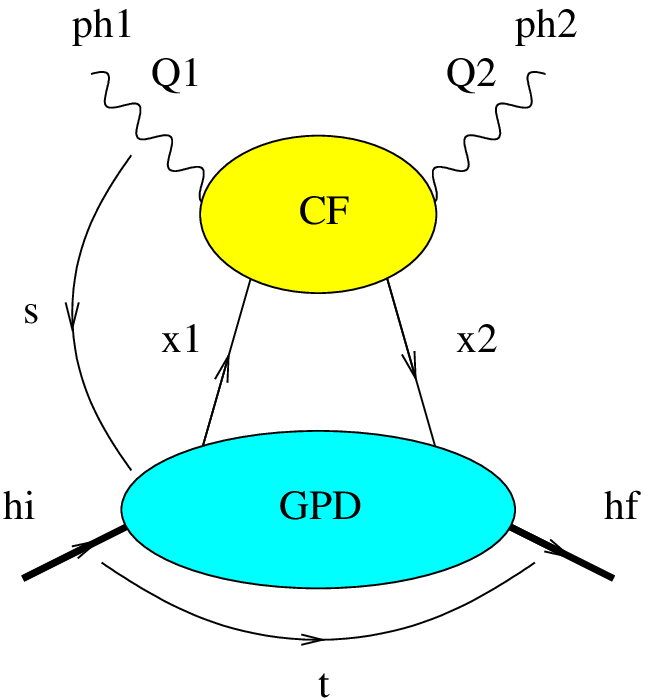}}}}
\end{tabular}}}
\caption{(a): DIS factorization.  (b): DVCS [TCS] factorization.}
\label{Fig:factDIS-DVCS}
\end{figure}

Historically, the first insight into the partonic content of the nucleon was obtained based on the deep inelastic scattering (DIS). As any {\it inclusive} process, the study of the total DIS cross-section is made based on the optical theorem, which relates this total cross-section to the \structure{forward} ($t=0$) Compton amplitude (see Fig.~\ref{Fig:factDIS-DVCS}a).
 The structure functions can be factorized collinearly as a convolution of coefficient functions (CFs) with parton distribution functions (PDFs).
 
The {\it exclusive} deep virtual Compton scattering (DVCS)  
and time-like Compton scattering (TCS),
 in the limit 
$s_{\gamma^* p}, \, Q^2 \gg -t\,,$ can also be factorized,  
now at the amplitude level (see Fig.~\ref{Fig:factDIS-DVCS}b). It involves generalised parton distribution functions (GPDs) 
\cite{Mueller:1998fvRadyushkin:1996ndJi:1996ek} which extend the PDFs outside of the diagonal kinematical limit: the $t$ variable as well as the longitudinal momentum transfer may not vanish, calling for new variables, the skewness $\xi$,
encoding the inbalance of longitudinal
$t-$channel momentum, and the 
transferred transverse momentum $\Delta$.
\psfrag{g}{$\gamma$}
\psfrag{gs}{$\gamma^*$}
\psfrag{s}{$s$}
\psfrag{t}{$t$}
\psfrag{Q2}{$\!\!\!Q^2$}
\psfrag{CF}{\!\!\reduced  CF}
\begin{figure}[h!]
\centerline{\scalebox{.77}{\centerline{\begin{tabular}{cc}
\psfrag{ph1}{$\gamma^*$}
\psfrag{s}{$s$}
\psfrag{t}{$t$}
\psfrag{hi}{$h$}
\psfrag{hf}{$h'$}
\psfrag{Q1}{$Q^2$}
\psfrag{GPD}{\raisebox{-.05cm}{\hspace{-.1cm}GPD}}
\psfrag{CF}{\raisebox{-.05cm}{\hspace{-.1cm}CF}}
\psfrag{DA}{\raisebox{.15cm}{\rotatebox{-55}{DA}}}
\psfrag{x1}{\hspace{-.4cm}$x+\xi$}
\psfrag{x2}{$x-\xi$}
\psfrag{z}{$u$}
\psfrag{zb}{$-\bar{u}$}
\psfrag{rho}{$\rho, \, \pi$}
\hspace{-.5cm}
\raisebox{-.44 \totalheight}{\includegraphics[height=4.8cm]{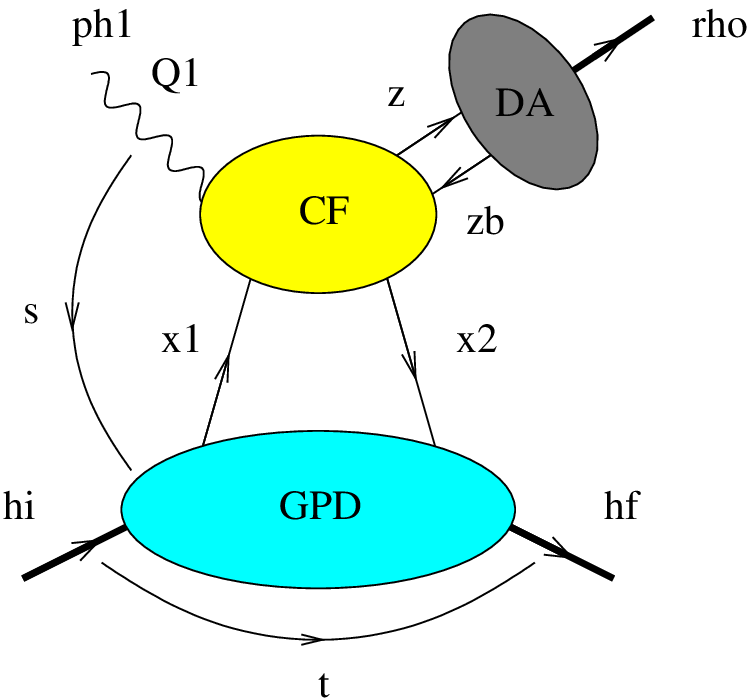}}
&
\psfrag{h}{ hadron}
\psfrag{GDA}{\raisebox{-.05cm}{\hspace{-.12cm}GDA}}
\psfrag{CF}{\raisebox{-.05cm}{\hspace{-.1cm}CF}}
\hspace{1cm}\raisebox{-.4 \totalheight}{\includegraphics[height=4.5cm]{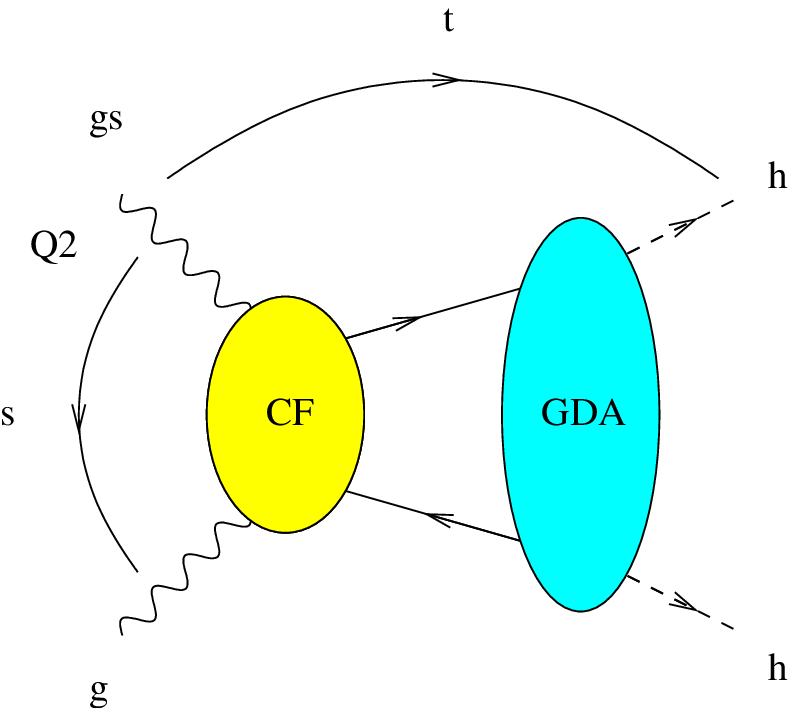}}
\end{tabular}}}}
\caption{(a): Collinear factorization of meson electroproduction.
 (b): Collinear factorization of hadron pair production in $\gamma \gamma^*$ subchannel.}
\label{Fig:factRho-double_meson}
\end{figure}

From DVCS, several extensions have been made. First, one may replace the produced $\gamma$ by a meson, factorized collinearly through a DA~\cite{Collins:1996fbRadyushkin:1996ru} (Fig.~\ref{Fig:factRho-double_meson}a).  Second, one may consider the crossed process 
in the limit $s_{\gamma^* p}  \ll -t, \,Q^2$. It again factorizes (Fig.~\ref{Fig:factRho-double_meson}b), the $q \, \bar{q}$ content of the hadron pair being encoded in a generalised distribution amplitude (GDA)~\cite{Diehl:1998dk}. 
To illustrate how the collinear factorization sets in depending on the kinematical regime, one may consider the
process $\gamma^*(q) \, \gamma \to \gamma \, \gamma\,.$
At large $-q^2$ and in the large center-of-mass energy limit, it factorizes in terms of the photon GPD, while at large $-q^2$ and in the threshold limit, it factorizes in terms of the diphoton GDA~\cite{Friot:2006mmElBeiyad:2008ss}. 
These frameworks allow to describe 
hard exotic hybrid meson production both in electroproduction  and $\gamma \gamma^*$ collisions (including its decay mode, e.g. $\pi \, \eta$)~\cite{Anikin:2004vcAnikin:2004jaAnikin:2006du}.

\begin{figure}
\centerline{
\scalebox{.8}{\begin{tabular}{ccc}
\psfrag{ph1}{$\gamma^*$}
\psfrag{ph2}{$\gamma$}
\psfrag{s}{$s$}
\psfrag{t}{$t$}
\psfrag{hi}{$h$}
\psfrag{hf}{$h'$}
\psfrag{Q1}{$Q^2$}
\psfrag{Q2}{}
\psfrag{GPD}{\raisebox{-.05cm}{\hspace{-.1cm}GPD}}
\psfrag{CF}{\raisebox{-.05cm}{\hspace{-.1cm}CF}}
\psfrag{x1}{\hspace{-.4cm}$x+\xi$}
\psfrag{x2}{$x-\xi$}
\raisebox{-.44 \totalheight}{\includegraphics[height=5cm]{dvcs.eps}}
&
\scalebox{1.3}{$\stackrel{t \, \to \, u}{\longrightarrow}$}
&
\psfrag{hi}{$h$}
\psfrag{hf}{$h'$}
\psfrag{Q1}{$Q^2$}
\psfrag{Q2}{}
\psfrag{s}{$s$}
\psfrag{t}{$t$}
\psfrag{ph1}{$\gamma^*$}
\psfrag{ph2}{$\gamma$}
\psfrag{x1}{\hspace{-.4cm}$x+\xi$}
\psfrag{x2}{$x-\xi$}
\psfrag{GPD}{\raisebox{-.05cm}{\hspace{-.1cm}TDA}}
\psfrag{CF}{\raisebox{-.05cm}{\hspace{-.1cm}CF}}
\raisebox{-.44 \totalheight}{\includegraphics[height=5cm]{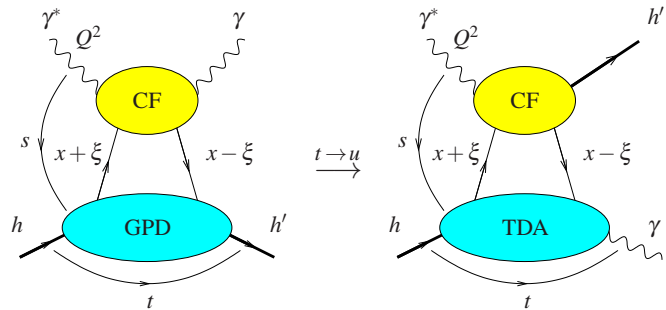}}
\end{tabular}}}
\caption{$t \leftrightarrow u$ crossing from DVCS.
}
\label{Fig:factTDA}
\end{figure}

Starting from usual DVCS, the next extension is to allow the  initial and the final hadron to differ (in the same $SU(3)$ octuplet), replacing GPDs by transition GPDs. 
To be even less diagonal, the conservation of the baryonic number can be removed between inital and final state, introducing transition distribution amplitudes (TDAs)~\cite{Pire:2004ie}. This can be obtained from 
DVCS by a $t \leftrightarrow u$ crossing,  as shown in Fig.~\ref{Fig:factTDA}. 
%
\psfrag{p}{$p$}  
\psfrag{pp}{$p'$}
\psfrag{q}{$q$}
\psfrag{qp}{$q'$}
\psfrag{pip}{$p$}
\psfrag{pim}{$\pi^-$}
\psfrag{pr}{$p$}
\psfrag{apr}{$\bar p$}
\psfrag{g}{$\gamma$}
\psfrag{gs}{$\gamma^*$}
\psfrag{u}{$u$}
\psfrag{db}{$\bar d$}
\psfrag{ep}{$e^+$}
\psfrag{em}{$e^-$}
\psfrag{d}{$d$} 
\psfrag{a}{$a$}
\psfrag{b}{$b$}
\psfrag{pi}{$\pi$}
\psfrag{k}{$k$}
\psfrag{DA}{$\,\,DA$}
\psfrag{TDA}{$TDA$}
\psfrag{TH}{$T_H$}
\psfrag{q}{}
\psfrag{pim}{}
A further extension is done by replacing the outoing \alert{$\gamma$} by any hadronic state~\cite{Pire:2005axLansberg:2007se}. In particular,
the $p \to \gamma$ and 
$p \to \pi$
TDAs could be measured in the forward scattering of a $\bar{p}$ beam on a $p$ probe, as planned by the PANDA  collaboration at GSI-FAIR~\cite{Lansberg:2007seLansberg:2012ha}.

As a theoretical playground, the process $\gamma^* \, \gamma^* \to \rho^0_L \, \rho^0_L$ is of particular interest. Indeed, depending on the polarization of the incoming photons, it  can be factorized in two ways involving either the GDA of the $\rho$ pair (for $\gamma^*_T$) or the $\gamma^* \to \rho$
TDA (for $\gamma^*_L$)~\cite{Pire:2006ik}.


\subsection{The twist-2 GPDs}

\begin{figure}[h!]
\begin{center}
     \leavevmode
     \epsfxsize=.91\textwidth

     \epsffile{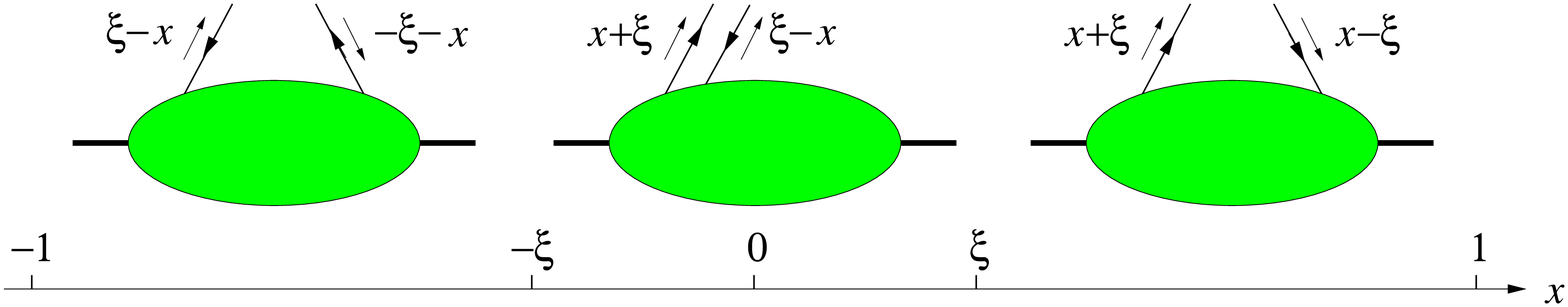}
\end{center}
\begin{tabular}{ccc}
\hspace{0.5cm}
\begin{tabular}{c}
\structure{\footnotesize Emission and reabsoption} \\
\structure{\footnotesize of an antiquark}\\
\structure{\footnotesize $\sim$ PDFs for antiquarks}\\
\quad {\aut DGLAP}-II region
\end{tabular}
&
\hspace{-.6cm}
\begin{tabular}{c}
\structure{\footnotesize Emission of a quark and} \\
\structure\footnotesize {emission of an antiquark}\\
\structure{\footnotesize $\sim$ meson exchange}\\
\quad {\aut ERBL} region
\end{tabular}
&
\hspace{-.6cm}
\begin{tabular}{c}
\structure{\footnotesize Emission and reabsoption} \\
\structure{\footnotesize of a quark}\\
\structure{\footnotesize $\sim$ PDFs for quarks}\\
\quad {\aut DGLAP}-I region
\end{tabular}
\end{tabular}
\caption{\label{Fig:regions_GPD} The parton interpretation of GPDs in
 the three $x$-intervals. Figure from~\cite{Diehl:2003ny}.
}
\end{figure}

The twist 2 GPDs have a simple
physical interpretation, shown in Fig.~\ref{Fig:regions_GPD}. Their
classification goes as follows, according to the fact that 
the considered non-perturbative matrix elements $F^q$ and $\tilde{F}^q$ are diagonal in helicity or not.

\bei

\item For (massless) quarks, this can be equivalently formulated in terms of the chirality of the $\Gamma$ matrix involved in the bilocal light-cone operators which matrix element define $F^q$ and $\tilde{F}^q$.  One should distinguish the exchanges
	\bei
	\item \structure{without helicity flip} (\structure{chiral-even} \alert{$\Gamma$} matrices), \alert{4 chiral-even GPDs} :

\alert{$H^q$} ($\xrightarrow{\tiny\xi=0,\, t=0}$ \structure{PDF} $q$) \alert{, $E^q$, $\tilde{H}^q$} $(\xrightarrow{\tiny\xi=0,\,t=0}$ \structure{polarized PDFs} $\Delta q$) and \alert{$\tilde{E}^q$},
\beqas
  \label{GPD_quark}
&&\hspace{-2.2cm} F^q =
\frac{1}{2} \int \frac{d z^+}{2\pi}\, e^{ix P^- z^+}
  \langle p'|\, \bar{q}(-\half z)\, {\alert{\gamma^-}} q(\half z) 
  \,|p \rangle \Big|_{z^-=0,\,\, \tvec{z}=0}
\nonumber \\
&&\hspace{-2.2cm} \phantom{F^q} = \frac{1}{2P^-} \left[
  \alert{H^q}(x,\xi,t)\, \bar{u}(p') \gamma^- u(p) +
  \alert{E^q}(x,\xi,t)\, \bar{u}(p') 
                 \frac{i \,\sigma^{-\alpha} \Delta_\alpha}{2m} u(p)
  \, \right] ,
\nonumber \\
&&\hspace{-2.2cm}\tilde{F}^q =
\frac{1}{2} \int \frac{d z^+}{2\pi}\, e^{ix P^- z^+}
  \langle p'|\, 
     \bar{q}(-\half z)\, \alert{\gamma^- \gamma_5}\, q(\half z)
  \,|p \rangle \Big|_{z^-=0,\, \,\tvec{z}=0}
\nonumber \\
&&\hspace{-2.2cm} \phantom{F^q} = \frac{1}{2P^-} \left[
 \alert{\tilde{H}^q}(x,\xi,t)\, \bar{u}(p') \gamma^- \gamma_5 u(p) +
  \alert{\tilde{E}^q}(x,\xi,t)\, \bar{u}(p') \frac{\gamma_5 \,\Delta^-}{2m} u(p)
  \, \right] .
\eqas

	\item
	\structure{with helicity flip} ( \structure{chiral-odd} \alert{$\Gamma$}  mat.), 
\alert{4 chiral-odd GPDs:} 

$H^q_T$ ($\xrightarrow{\tiny\xi=0,\,t=0}$ \structure{quark transversity PDFs} $\Delta_T q$\alert{), $E^q_T$, $\tilde{H}^q_T$, $\tilde{E}^q_T$}
\beqas
&&\hspace{-2.2cm}\frac{1}{2} \int \frac{d z^+}{2\pi}\, e^{ix P^- z^+}
  \langle p'|\, 
     \bar{q}(-\half z)\, i \, \alert{\sigma^{-i}}\, q(\half z)\, 
  \,|p \rangle \Big|_{z^-=0,\,\, \tvec{z}=0} 
 \\
&&\hspace{-2.21cm}= \!\frac{1}{2P^-} \bar{u}(p') \! \!\!\left[\!
 \alert{H_T^q}\, i \sigma^{-i} \!\!\!+\!
  \alert{\tilde{H}_T^q}\, \frac{P^- \Delta^i - \Delta^- P^i}{m^2} \!+\!
  \alert{E_T^q}\, \frac{\gamma^- \Delta^i - \Delta^- \gamma^i}{2m} \!+\!
  \alert{\tilde{E}_T^q} \frac{\gamma^- P^i - P^- \gamma^i}{m}
  \right] \!\! u(p)  , \nonumber
\eqas
	\ei

\item A similar analysis can be made for twist-2 gluonic GPDs (for which the notion of chirality does not make sense):

	\bei

	\item \alert{4 gluonic GPDs} \structure{without helicity flip:} 

\alert{$H^g$} ($\xrightarrow{\tiny\xi=0,\,t=0}$ \structure{PDF} $x \, g$), 
\alert{$E^g$},
\alert{$\tilde{H}^g$} ($\xrightarrow{\tiny\xi=0,\,t=0}$ \structure{polarized PDF} $x \, \Delta g$) and
\alert{$\tilde{E}^g$}

	\item \alert{4 gluonic GPDs} \structure{with helicity flip:} 
\alert{$H^g_T$},
\alert{$E^g_T$},
\alert{$\tilde{H}^g_T$} and 
\alert{$\tilde{E}^g_T$}.
We note that there is 
no forward limit reducing to gluon PDFs here: a change of  2 units of helicity cannot be  compensated by a spin 1/2 target.
	\ei

\ei


%
\subsection{Transversity}

The extraction of GPDs from DVCS measurements will soon enter a precision era, in particular with the expected bunch of data provided by JLab@12GeV and COMPASS-II. One should however note that this concerns only the quark and gluon GPDs with a zero total helicity transfer. The sector of GPDs involving a non zero helicity transfer introduced in the previous subsection is almost completely unknown. The tranverse spin content of the proton is related to non-diagonal helicity observables, since
\beqas
\hbox{spin along }  x : \quad
\begin{tabular}{ccc}$| \uparrow \rangle_{(x)}  $  & $\sim$ & $| \rightarrow\rangle +| \leftarrow\rangle$\\
$| \downarrow \rangle_{(x)}  $ & $\sim$ & $| \rightarrow\rangle -| \leftarrow\rangle$ 
\end{tabular} \quad \hbox{: helicity states}
\,.
\eqas
An observable sensitive to helicity spin flip
gives thus access to the transversity PDF
\alert{$\Delta_T q(x)$},
which is very badly known. Since for massless (anti)quarks chirality = (-) helicity,  
transversity is a chiral-odd quantity. This implies that transversity cannot be extracted from usual fully inclusive DIS. Based on the fact that chirality must be flipped twice, one can consider processes with either  two hadrons
in the initial state, like in proton-proton collision or one hadron in the initial state with at least
one hadron in the final state, as in the case of semi-inclusive DIS.

Let us move from the inclusive to the exclusive case.
Again, 
since 
transversity is a chiral-odd quantity, and based on the fact that QCD and QED are chiral even in the massless limit, any chiral-odd operator should be balanced by another chiral-odd operator in the amplitude of any exclusive process.
\begin{figure}[h!]
\psfrag{z}{\hspace{-0.1cm}\footnotesize $z$ }
\psfrag{zb}{\raisebox{0cm}{\hspace{-0.1cm} \footnotesize$\bar{z}$} }
\psfrag{gamma}{\raisebox{+.1cm}{\footnotesize $\,\gamma$} }
\psfrag{pi}{\footnotesize$\!\pi$}
\psfrag{rho}{\footnotesize$\,\rho$}
\psfrag{TH}{\raisebox{-.05cm}{\hspace{-0.1cm}\footnotesize $T_H$}}
\psfrag{tp}{\raisebox{.3cm}{\footnotesize $t'$}}
\psfrag{s}{\hspace{0.4cm}\footnotesize$s$ }
\psfrag{Phi}{}
\vspace{.3cm}
\scalebox{1.2}{\hspace{-1.3cm}\centerline{\hspace{-2.7cm} 
\scalebox{.9}{\raisebox{.6cm}{\includegraphics[width=4.3cm]{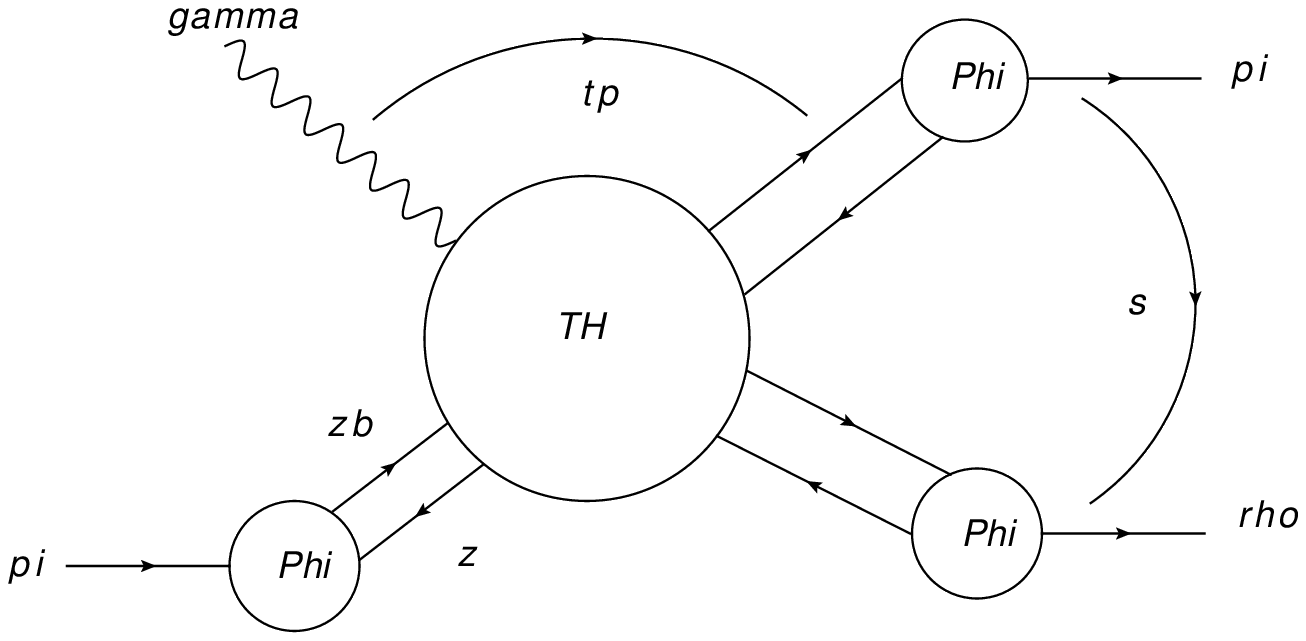}}}\hspace{0.5cm} \raisebox{1.5cm}{$\longrightarrow$}\hspace{0.2cm}
\psfrag{piplus}{\footnotesize$\,\pi^+ $ chiral-\alert{even} twist 2 DA}
\psfrag{rhoT}{\footnotesize$\,\rho^0_T$ chiral-\alert{odd} twist 2 DA}
\psfrag{M}{\hspace{-0.15cm} \footnotesize $M^2_{\pi \rho}$ }
\psfrag{x1}{\raisebox{-.1cm}{\hspace{-0.6cm} \footnotesize $x+\xi $  }}
\psfrag{x2}{\raisebox{-.1cm}{\hspace{-0.1cm} \footnotesize  $x-\xi $ }}
\psfrag{N}{ \hspace{-0.4cm}\footnotesize $N$}
\psfrag{GPD}{\raisebox{-.1cm}{\footnotesize \hspace{-0.4cm}  $GPDs$}}
\psfrag{Np}{\footnotesize$N'$}
\psfrag{t}{ \raisebox{-.1cm}{\footnotesize \hspace{-0.3cm}   $t \ll M^2_{\pi \rho}$  \hspace{.1cm}\raisebox{-.1cm}{chiral-\alert{odd} twist 2 GPD}}}
\scalebox{.9}{\includegraphics[width=4.5cm]{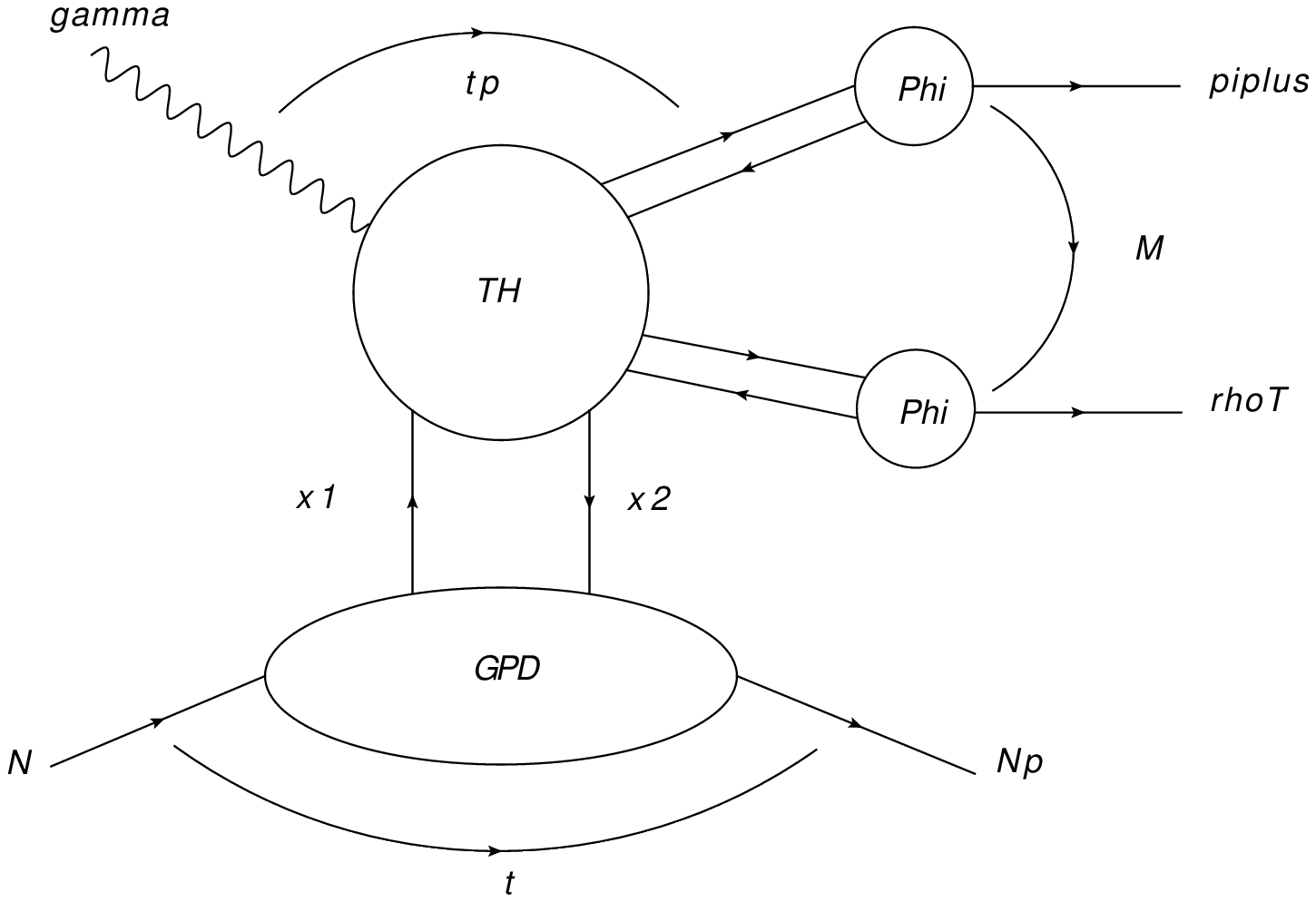}}}}
\vspace{0cm}
\caption{Brodsky-Lepage factorization applied to $\gamma \, N \to \pi^+ \, \rho^0_T \, N'$.}
\label{Fig:transversity3}
\end{figure}
Since the dominant DA for $\rho_T$ is of twist 2 and chiral-odd, it seems natural to consider $\rho_T$-electroproduction.
Unfortunately
the amplitude vanishes,
at any order in perturbation theory, since this process would require a transfer of 2 units of helicity from the proton~\cite{Diehl:1998pdCollins:1999un}.
Although this vanishing is true only at twist 2,
 processes involving twist 3 DAs~\cite{Ball:1998skBall:1998ffBall:2007zt} may face problems with factorization
(see Sec.~\ref{SubSec:PB}).
One can circumvent this vanishing by considering a 3-body final state~\cite{Ivanov:2002jjEnberg:2006heBeiyad:2010cxa}.
Indeed the process
$\gamma \, N \to \pi^+ \, \rho^0_T \, N'$ can be described in the spirit of large angle factorization~\cite{Brodsky:1981rp}
of the process
$\gamma \, \pi \rightarrow \pi \, \rho $ \structure{at large $s$ and fixed angle} (i.e. for fixed $t'/s, \, u'/s$ in Fig.~\ref{Fig:transversity3}), $M_{\pi\rho}^2$ providing the hard scale, as considered in sec.~\ref{subsec:early}. Besides its interest for the transversity sector, one should note that
such processes with a 3-body final state can give access to all GPDs,
$M_{\pi\rho}^2$ playing the role of the $\gamma^*$ virtuality of usual TCS. 

On the theoretical side, it is also of particular importance to build a consistent framework for any modeling of the transversity quark and gluon GPDs. This can be achieved 
based on a double partial wave expansion (in the conformal and $SO(3)$ partial waves) in the cross channel. Equivalently, this general formulation can be obtained by an explicit calculation
of the cross channel spin-$J$ resonance exchange contributions~\cite{Pire:2014fwa}.



\subsection{Resummation effects}

\begin{figure}[h]
\centerline{\scalebox{1}{
\begin{tabular}{c}
\hspace{.25cm}\raisebox{-.44 \totalheight}{
\psfrag{p}[cc][cc]{}
\psfrag{q1}[cc][cc]{$q$}
\psfrag{gas}[cc][cc]{\raisebox{1cm}{$\hspace{-.5cm}\gamma^*$}}
\psfrag{g}[cc][cc]{\raisebox{1cm}{$\hspace{.1cm}\gamma$}}
\psfrag{H}[cc][cc]{$T^q$}
\psfrag{k1}[cc][Bc]{\raisebox{0cm}{$\hspace{-.88cm} (x+\xi) p$}}
\psfrag{k2}[cc][Bc]{\raisebox{0cm}{$\hspace{.68cm} (x-\xi) p$}}
\psfrag{S}[cc][cc]{${\cal S}$}
\psfrag{U}[cc][cc]{${\cal U}$}
\includegraphics[height=3.375cm]{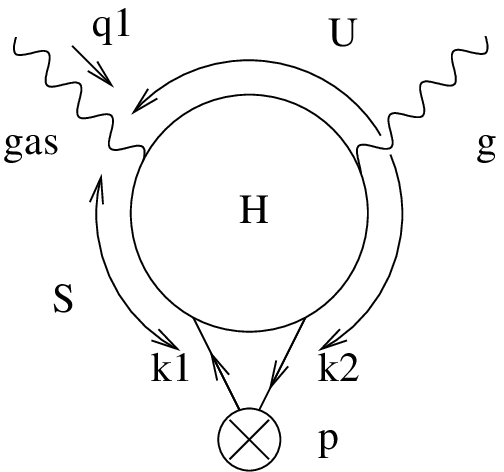}} 
\\ 
%
\psfrag{pl}[rc][Br]{}
\psfrag{pr}[cc][Bc]{}
\psfrag{pp}[rc][rc]{}
\psfrag{ppp}[cc][lc]{}
\psfrag{pmq}[cc][cc]{$+ \, -$}
\psfrag{mu}[cc][cc]{$+$}
\psfrag{n}[cc][cc]{}
\psfrag{S}[cc][cc]{$F^q$}
\hspace{.57cm}
\raisebox{-1 \totalheight}{\includegraphics[height=2.7cm]{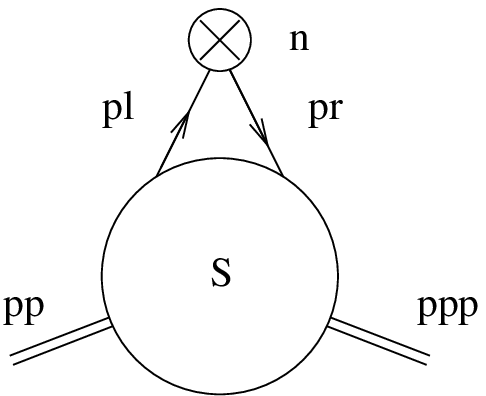}} 
\end{tabular}
}}
\caption{Factorization of the DVCS amplitude in the hard regime. The crossed-blob denote an appropriate set of Dirac $\Gamma$ matrices.}
\label{Fig:factGPD}
\end{figure}

Consider the usual collinear 
factorization of the DVCS amplitude as a convolution of  coefficient functions with GPDs (see Fig.~\ref{Fig:factGPD} for the quark case).
The DVCS coefficient function has  threshold singularities in its $s-$ and $u$-channels, in the limits $x \to \pm \xi\,.$
Indeed, considering the invariants ${\cal S}$ and ${\cal U}$
for the coefficient function,
\beqas
{\cal S}= \phantom{-}\frac{x-\xi}{2 \xi} \, Q^2  &\ll& Q^2 \quad \hbox{ when } x \to \xi \\  {\cal U}= - \frac{x+\xi}{2 \xi} \, Q^2   &\ll& Q^2 \quad \hbox{ when } x \to -\xi
\label{def-S-U}
\eqas
which means that one pass from a single-scale analysis w.r.t. $Q^2$ to a two scales problem, a typical situation which calls for threshold singularities to be resummed.
It turns out that 
soft-collinear effects lead to large terms of type
$[\alpha_S \log^2(\xi\pm x)]^n/(x\pm \xi)$ which can be resummed in light-like gauge as ladder-like diagrams~\cite{Altinoluk:2012ntAltinoluk:2012fb}.

\subsection{Limitations within the collinear factorization framework}
\label{SubSec:PB}

The collinear factorization  is known to be applicable for a limited number of cases. Consider for example the case of $\rho-$electroproduction.
Since QED and QCD vertices are chiral even (in the massless limit),
the total helicity of a $q \bar{q}$ pair produced by a $\gamma^*$ should vanish, and the  $\gamma^*$ helicity equals the $q \bar{q}$ orbital momentum $L^{q \bar{q}}_z$. In the pure collinear limit (i.e. at twist 2), $L^{q \bar{q}}_z=0$, and thus the $\gamma^*$ is longitudinally polarized.
At $t=0$ there is no source of orbital momentum from the proton coupling 
so that the meson and photon helicities are identical. This statement is not modified in
the collinear factorization approach at $t \neq 0$. Indeed in collinear factorization the hard part should be treated as $t-$independent, since any $t$ dependency is power suppressed, i.e. should be considered as higher twist. This $s-$channel helicity conservation (SCHC) implies that the only allowed transitions are 
 $\gamma^*_L \to \rho_L$, for which QCD factorization  holds at twist 2 at any order in perturbation~\cite{Collins:1996fbRadyushkin:1996ru}, and 
$\gamma^*_T \to \rho_T$, for which QCD factorization faces
problems due to end-point singularities at twist 3 when integrating over quark longitudinal momenta~\cite{Mankiewicz:1999tt}.
The improved collinear approximation may be a solution: in this approach, one
 keeps a transverse $\ell_\perp$ dependency in the $q,$ $\bar{q}$ momenta,  to regulate end-point singularities.
 Now,
 soft and collinear gluon exchange between the valence quark are responsible for large double-logarithmic effects which are conjectured to exponentiate in a Sudakov factor~\cite{Li:1992nu}, regularizing
end-point singularities. This tail can be combined with an ad-hoc non-perturbative gaussian ansatz for the DAs, providing 
 practical tools for meson electroproduction phenomenology~\cite{Goloskokov:2005sdGoloskokov:2006hrGoloskokov:2007nt}.



\section{QCD at large $s$}

\subsection{Theoretical motivations}

The understanding of strong interaction in the Regge limit is a very fundamental question, which can be addressed based on perturbative methods.
The perturbative {\aut Regge} limit of QCD 
is reached 
in the diffusion of two hadrons $h_1$ and $h_2$ whenever
$\alert{\sqrt{s_{h_1 \, h_2}}}
 \gg$  other scales (masses, transfered momenta, ...), while 
other scales are assumed to be comparable (virtualities, etc...) and at least one of them is large enough to justify the applicability of perturbative QCD (photon virtuality, heavy final state, large $t$-channel exchanged momentum, large transverse momenta of the produced states, etc.). By inspection, one can show that loop corrections at large $s$ involve powers of $\ln s$, which might compensate the smallness of $\alpha_s$ which powers appear in these loops. This thus calls for a resummed approach. As a major step forward,
the dominant sub-series $\sum_n (\alpha_s \, \ln s)^n$
was computed in the middle of '70s, leading to
$\alert{\sigma_{tot}^{h_1\, h_2}}  \sim \alert{ s^{\alpha_\pom(0) -1}}$
($\alpha_\pom(0) >1$)~\cite{Fadin:1975cbKuraev:1976geKuraev:1977fsBalitsky:1978ic}, the so-called BFKL $\mathbb{P}$omeron
which violates  QCD $S$ matrix \alert{unitarity}. One of the
main issue of QCD is to improve this result, and to test this dynamics experimentally. The underlying high-energy QCD dynamics has been studied extensively in inclusive and semi-inclusive processes~\cite{test_inclusiveANDtest_semi_inclusive}. Based on existing (LHC) and forecoming facilities (LHeC, ILC) which combine both a large center-of-mass energy and a large luminosity, this can be studied in the even more challenging context of exclusive processes.


\subsection{$k_T$-factorization}

\begin{figure}
\psfrag{g1}[cc][cc]{$\gamma^*(q_1)$}
\psfrag{g2}[cc][cc]{$\gamma^*(q_2)$}
\psfrag{p1}[cc][cc]{\raisebox{.3cm}{$\,\rho(p_1)$}}
\psfrag{p2}[cc][cc]{\raisebox{.3cm}{$\,\rho(p_2)$}}

\psfrag{l1}[cc][cc]{}
\psfrag{l1p}[cc][cc]{}
\psfrag{l2}[cc][cc]{}
\psfrag{l2p}[cc][cc]{}
\psfrag{ai}[cc][cc]{$\stmath \beta^{\,\nearrow}$}
\psfrag{bd}[cc][cc]{$\stmath \alpha_{\,\searrow}$}
\psfrag{k}[cc][cc]{$k$}
\psfrag{rmk}[lc][cc]{$\hspace{-.5cm}r-k \hspace{3cm} \int d^2 k_\perp$}
\psfrag{oa}[cc][cc]{\scalebox{.9}{${}\quad {\tiny \stmath \alpha \ll \alpha_{\rm quarks}}$}}
\psfrag{ou}[ll][ll]{
\hspace{1cm} $\Rightarrow$ set $\alpha=0$ and  $\int d\beta$ }
\psfrag{ob}[cc][cc]{\scalebox{.9}{\raisebox{-1.5
    \totalheight}{$\quad {\tiny \stmath\beta \ll \beta_{\rm quarks}}$}}}
\psfrag{od}[ll][ll]{
\hspace{1cm} $\Rightarrow$ set $\beta=0$ and  $\int d\alpha$}
\scalebox{1}{
\centerline{\hspace{-3cm}\epsfig{file=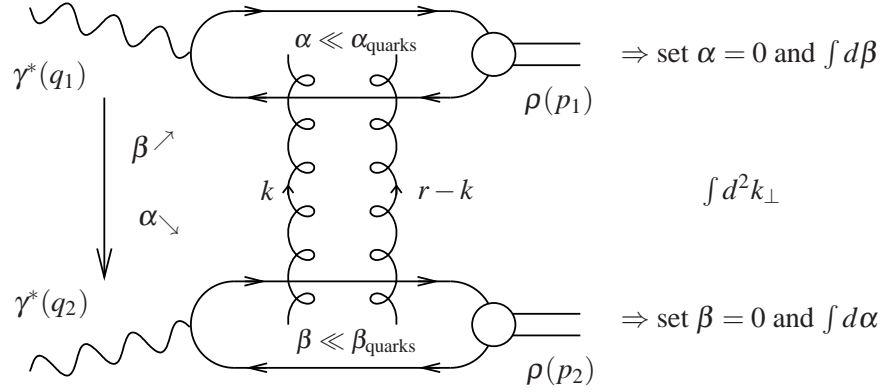,width=\widm}}
}
\caption{$k_T-$factorization applied to $\gamma^* \, \gamma^* \to \rho \, \rho\,.$}
\label{Fig:kT-factorization}
\end{figure}

The main tool in this regime is the $k_T$-factorization. Let us explain the main steps of this high-energy factorization, illustrated in Fig.~\ref{Fig:kT-factorization} for $\gamma^* \, \gamma^* \to \rho \, \rho$. First, introducing two light-like vectors $p_1$ and $p_2$ such that 
$2 p_1 \cdot p_2=s$ is a parametrically large scale, of the same order of magnitude as the squared center-of-mass energy
(in our example, one may chose $p_1$ and $p_2$ as the momenta of the two outgoing mesons),
it is convenient to 
use the {\aut Sudakov} decomposition of any 4-momentum as
$k = \alpha \,p_1 + \beta \, p_2 + k_\perp$, in which
${\stmath d^4k= \frac{s}{2} \, d \alpha \, d\beta \, d^2k_\perp}\,.$ 
In the large $s$ limit, keeping only the maximal powers of $s$, the numerator of any $t-$channel gluon can be written as a polarization sum over  the so-called \alert{non-sense} one, i.e.
$\varepsilon_{\alert{NS}}^{up}=\frac{2}s \, p_2$, $\varepsilon_{\alert{NS}}^{down}=\frac{2}s\,  p_1\,,$
since the momenta of the upper (lower) part of the diagram can be approximately considered as
flying along $p_1$ (resp. $p_2$), up to $s$ suppressed powers.
At high energy, the part of the phase space in the loop integral over $k$ which gives rise to a $\ln s$ corresponds to the approximation where  the $\alpha$ ($\beta$) component of $t$-channel gluons entering the upper (resp. lower) part of the diagram can be neglected. This simplifies considerably the $k$ integration, from which one obtains the
impact representation for exclusive processes amplitude\footnote{$\kb$ = Eucl. $\leftrightarrow $ $k_\perp$ = Mink.}
\beq
{\cal M} = is\;\int\;\frac{\stmath{d^2\,\kb}}{(2\pi)^2\stmath{\kb^2\,(\rb -\kb)^2}}
\alert{{\Phi}^{\gamma^*(q_1) \to \rho(p^\rho_1)}}(\stmath{\kb,\rb -\kb})\;
\alert{{\Phi}^{\gamma^*(q_2) \to \rho(p^\rho_2)}}(\stmath{-\kb,-\rb +\kb})\,,
\label{impact-rep}
\eq
where 
$\alert{{\Phi}^{\gamma^*(q_1) \to \rho(p^\rho_1)}}$ is the  $\gamma^*_{L,T}(q) g(k_1) \to \rho_{L,T}\, g(k_2)$ impact factor. One should note that for the upper (lower) part of the diagram,  $\beta$ (resp. $\alpha$) is proportional to the $s-$Mandelstam variable in the $\gamma^* g$ channel. Since the impact factors are defined as integral over $\beta$ (resp. $\alpha$) of $S-$matrix elements, they can be equivalently considered as the $s-$channel discontinuity of these $S-$matrix elements after closing the $\beta$ (resp. $\alpha$) integral over the right-hand cut. 


\subsection{Meson production}

From factorization point of view,
the ''easy'' case  is  $J/\Psi$ production, which mass provides the required hard scale
\cite{Ryskin:1992uiFrankfurt:1997fjEnberg:2002zyIvanov:2004vd}.
Exclusive vector meson photoproduction at \alert{large $t$} (providing the hard scale)
is another example (which however faces problem with end-point singularities) for which HERA data seems to favor a BFKL picture~\cite{Ivanov:2000uqEnberg:2003jwPoludniowski:2003yk}.
Exclusive electroproduction of vector meson  
can also be described~\cite{Goloskokov:2005sdGoloskokov:2006hrGoloskokov:2007nt}
based on the improved collinear factorization (see sec.~\ref{SubSec:PB}) for the coupling with the meson DA and collinear factorization for the GPD coupling.

The process $\gamma^{(*)} \gamma^{(*)} \to \rho \, \rho$
is an example of a realistic 
exclusive test of the $\pom$omeron, as  
a subprocess of 
$e^- \, e^+ \to e^- \, e^+ \, \rho_L^0 \, \rho_L^0$ with double lepton tagging. This could be measured  at ILC which should provide the required  very large energy
$(\sqrt{s} \sim 500$ GeV) and 
luminosity ($ \simeq 125\hbox{~fb}^{-1}/\hbox{year}$), with the planned 
detectors designed to cover the \alert{very forward} region, close from the beampipe~\cite{Pire:2005icEnberg:2005eqSegond:2007fjIvanov:2005gnIvanov:2006gtCaporale:2007vs}.

\begin{figure}
\psfrag{rh}[cc][cc]{$\hspace{.5cm}\rho$}
\psfrag{gam}[cc][cc]{$\hspace{-1.2cm}\gamma^{(*)}_{T,\,L}$}
\psfrag{PI}[cc][cc]{\scalebox{.7}{$\stmath \!\!\Psi_i$}}
\psfrag{PF}[cc][cc]{\scalebox{.7}{${\aut \!\!\Psi_f}$}}
\psfrag{pi}[cc][cc]{$\! \! p$}
\psfrag{pf}[cc][cc]{$\! \! p$}
\psfrag{sig}[cc][cc]{\scalebox{1.2}{$\alert{\hat{\sigma}}$}}
\psfrag{y1}[cc][cc]{}
\psfrag{fl}[cc][cc]{\scalebox{.85}{\raisebox{-0.35cm}{$\,\updownarrow \!x_{\perp}$}}}
\psfrag{P1}[cc][cc]{}
\psfrag{PA3}[cc][cc]{}
\centerline{\epsfig{file=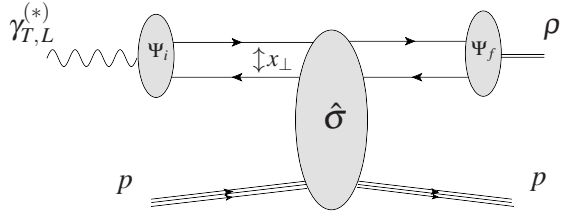,width=6.8cm}}
\caption{Dipole representation for $\gamma^* p \to \rho p$ high energy scattering.}
\label{Fig:dipole}
\end{figure}

Diffractive vector meson electroproduction have recently been described beyond leading twist, combining collinear factorization and $k_T-$factorization.
Based on the $\gamma^*_{L,T} \to \rho_{L,T}$
impact factor
including two- and three-partons contributions, one can describe HERA data on the ratio of the dominant helicity amplitudes~\cite{Anikin:2009hkAnikin:2009bfAnikin:2011sa}.
The dipole representation of high energy scattering~\cite{Mueller:1989stNikolaev:1990ja}
 (Fig.~\ref{Fig:dipole}), equivalent to the BFKL approach~\cite{Chen:1995paNavelet:1997tx},
is very convenient to implement saturation effects,
through a universal  proton-dipole  scattering amplitude $\hat{\sigma}(x_{\perp})$
~\cite{GolecBiernat:1998jsGolecBiernat:1999qd}.
Data for $\rho$ production call for models encoding saturation~\cite{Munier:2001nrKowalski:2006hc}. 
This dipole representation is consistent with the \structure{twist 2} collinear factorization, and it has been recently proven that it remains valid
beyond leading twist. This leads to a very good description of HERA data, except at low $Q^2$ where higher twist corrections seem to be rather important~\cite{Besse:2012iapheno_saturation}. An impact parameter analysis in this spirit would be very interesting, since it provides a probe of the proton shape, in particular through local geometrical scaling \cite{Ferreiro:2002kvMunier:2003bf}.

\subsection{Looking for the Odderon through exclusive processes}

The $\mathbb{O}$dderon hunting, the elusive $C-$odd partner of the $\mathbb{P}$omeron, has not been successful yet in any hard process, despite its predicted existence. Contrarily to the case of the $\mathbb{P}$omeron, which has an  intercept $\alpha_\pom(0)-1$ positive, the $\mathbb{O}$dderon is expected to have a vanishing intercept~\cite{Bartels:1999ytKovchegov:2003dm}.

Several strategies have been pursued in order to reveal it. First, one may consider exclusive 
processes where the $\mathcal{M}_\mathbb{P}$ amplitude vanishes due to $C$-parity conservation~\cite{Ginzburg:1992miBraunewell:2004pfBzdak:2007cz},
the signal being quadratic in the $\mathcal{M}_\mathbb{O}$ contribution. Second, one may
 consider observables sensitive to the \alert{interference} between $\mathcal{M}_\mathbb{P}$ and $\mathcal{M}_\mathbb{O},$ like asymmetries, thus providing observables \alert{linear} in $\mathcal{M}_\mathbb{O}$~\cite{Brodsky:1999mzGinzburg:2002zdGinzburg:2003ciHagler:2002nhHagler:2002sgHagler:2002nfPire:2008xe}.

\section{Conclusion}

Since a decade, there have been much progress in the understanding of hard exclusive processes.  
There is now a consistent framework starting from first principles, in order to deal with
medium energy exclusive processes, starting from DVCS. This allows to describe a huge number of processes. At high energy, the impact representation is a powerful tool for describing exclusive processes in diffractive experiments; they are and will be essential for studying QCD in the hard Regge limit ($\pom$omeron, $\odd$dderon, saturation...). 
Still, some problems remain: from the theory side,
proofs of factorization have been obtained only for a very few processes
(ex.: $\gamma^* \, p \to \gamma \, p\,$, $\gamma^*_L \, p \to \rho_L \, p$). For some other processes, it is highly plausible, but  not fully demonstrated, like those involving GDAs and TDAs. Furthermore, some processes explicitly show sign of breaking of factorization
(ex.:  $\gamma^*_T p \to \rho_T p$  at leading order), and a precise factorization scheme   starting from first principles is still missing in these situations.
The effect of QCD evolution, the NLO corrections and the choice of renormalization/factorization scale~\cite{Brodsky:1982gcAnikin:2004jbMoutarde:2013qs}, as well as power corrections (including a complete classification, which has been recently explored for scalar target~\cite{Pire:2013vea}) will be very relevant to interpret and describe the forecoming data, in particular in future facilities like EIC~\cite{Boer:2011fh} or LHeC~\cite{AbelleiraFernandez:2012cc}. A first principles description of the whole set of non-perturbative correlators occuring in exclusive processes is out of reach, since it would require to solve the confinement problem. A promising approach has been explored, based on the
AdS/QCD correspondence. This may provide insight for modeling the involved non-perturbative  correlators~\cite{Brodsky:2006uqaGao:2009seMarquet:2010sf}. However, the minimal version of the AdS/QCD correspondence does not seem to give results compatible with phenomenological constraints.

\end{document}